# High-Speed and Energy-Efficient Non-Volatile Silicon Photonic Memory Based on Heterogeneously Integrated Memresonator

Bassem Tossoun[1*], Di Liang[1], Stanley Cheung[1], Zhuoran Fang[1], Xia Sheng[1], John Paul Strachan[1] and Raymond G. Beausoleil[1]

[1]Hewlett Packard Labs, Hewlett Packard Enterprise, 820 N McCarthy Blvd, Milpitas, 95305, CA, USA.

*Corresponding author. Email: bassem.tossoun@hpe.com

**Recently, interest in programmable photonics integrated circuits has grown as a potential hardware framework for deep neural networks, quantum computing, and field programmable arrays (FPGAs). However, these circuits are constrained by the limited tuning speed and high power consumption of the phase shifters used. In this paper, introduced for the first time are memresonators, or memristors heterogeneously integrated with silicon photonic microring resonators, as phase shifters with non-volatile memory. These devices are capable of retention times of 12 hours, switching voltages lower than 5 V, an endurance of 1,000 switching cycles. Also, these memresonators have been switched using 300 ps long voltage pulses with a record low switching energy of 0.15 pJ. Furthermore, these memresonators are fabricated on a heterogeneous III-V/Si platform capable of integrating a rich family of active and passive optoelectronic devices directly on-chip to enable in-memory photonic computing and further advance the scalability of integrated photonic processors.**

Over recent years, the demand for high performance computers (HPCs) capable of efficiently running artificial intelligence applications has grown dramatically. The number of programs which use deep learning training has doubled every 3.5 months, which is much faster than the rate of performance doubling predicted by Moore's law [1]. In addition, learning algorithms are required to be executed in real-time on a massive amount of data produced by the plethora of interconnected smart devices within the Internet of Things (IoT) and edge computing.

Today, AI algorithms utilized by applications such as autonomous driving vehicles and Amazon's Alexa, are implemented using neural networks (NNs), a model inspired by the neuro-synaptic network within the human brain, which is the most energy-efficient computer to human knowledge (able to process 10 petaflops of data with only 20 W of power) [2]. The most commonly used hardware for running NNs includes application specific integrated circuits (ASICs), graphics processing units (GPUs) and field-programmable gate arrays (FPGAs). Current state-of-the-art electronic accelerators consume about 0.5 pJ in processing a single multiply-accumulate (MAC) operation, the most fundamental neural network calculation [3].

While conventional microelectronic processor performance progressed in line with Moore's Law as transistor density increased and multi-core processors developed, they are still fundamentally limited in both speed and power. Joule heating and the charging of metal wires involved in the movement of data constrain the operating speed and dominate the power consumption within electronic neural network hardware [4]. To exacerbate this issue even further, the von Neumann bottleneck and the "memory wall" restrict the bandwidth of data communications between the processor and the memory. Furthermore, digital processing units are also bottlenecked by the clock rate of the processor and their ability to efficiently compute single multiply-accumulate (MAC) operations, the most fundamental neural network calculation.

Fortunately, several technological breakthroughs over the last few decades have opened novel opportunities to battle these challenges. Silicon photonics offers a promising solution to dramatically improve the bandwidth and energy-efficiency of interconnects for data communications applications including data centres and HPCs [5]. Most recently, silicon photonics has not only been used for data communications, but for non-von Neumann accelerators used for applications such as deep learning [6, 7, 8, 9, 10, 11]. Some of the inherent properties of photonics make it a suitable platform for neuromorphic computing such as its high bandwidth of data transmission and parallel operation enabled by unique multiplexing schemes like wavelength division multiplexing (WDM). Furthermore, the processing time scale of a photonic neuron is within femtoseconds, which is orders of magnitude higher than electronic counterparts [12].

Because running a task on a deep neural network often can take a significant amount of time, there is a significant benefit to having non-volatile memory on-chip as it eliminates the static power consumption in holding weight values throughout an inference task. On-chip memory also prevents the need to retrieve results stored on a separate memory chip in between epochs or training steps. In addition, non-volatile photonic memory is not only useful for data storage, but also as part of the computational algorithms running on photonic neuromorphic computers [4]. More specifically, high-speed and low-power non-volatile photonic phase shifters are essential in enabling a larger variety of machine learning methods to be executed on integrated optical neural networks. For example, deep neural networks utilizing online training with algorithms such as backpropagation require synaptic weights to be updated frequently. These on-the-fly learning algorithms are scalable, memory-efficient, and can even be used to circumvent the losses compounded by the device imperfections within photonic neural networks as they scale in size and complexity [13, 14].

One viable solution to supplying a fast, low-power, non-volatile memory is the memristor (also commonly referred to as resistive random-access memory or RRAM) which was theoretically proposed by Leon Chua and experimentally demonstrated by HP Labs [15, 16].

Memristors (also commonly referred to as resistive random-access memory or RRAM) have proven to be excellent non-volatile electronic memory devices with high switching speed (~100 ps), low energy switching (~100 fJ), endurance ($10^{12}$ cycles), and high density [17, 18, 19, 20, 21].

In this work, we integrated metal oxide-based memristive devices within III-V/Si microring resonators to produce memresonators, an energy-efficient analogue non-volatile memory on a highly scalable and versatile heterogeneous silicon photonic platform well-suited for integrated photonic information processing circuits. By changing the resistance state of the memristor, we can subsequently tune the optical phase within the waveguide and alter the resonant wavelength of the device. Analogue device operation was shown through the measurement of multiple optical states. Performance records including retention times of 12 hours, an endurance of 1,000 switching cycles, switching times as low as 300 ps for SET and 900 ps for RESET, and switching energies of 0.15 pJ for SET and 0.36 pJ for RESET are demonstrated.

By integrating these memristors on the same chip as photonic neural networks, for example, significant amounts of energy and latency can be saved by avoiding energy lost in the transfer of data from the processor to an external memory chip. Moreover, using these memresonators, weights within photonic neural networks can be stored and updated at high speeds and low energy, enabling the use of the back-propagation algorithm and the ability to train the network on-chip. Finally, this III-V-on-silicon photonic memristive device is based on the same technology developed for a fully active (including optical gain) and passive integrated photonic platform on silicon for large-bandwidth, energy-efficient optical interconnect applications [22]. In fact, the first generation of a heterogeneous III-V-on-silicon technology has been successfully commercialized by Intel in their 300 mm CMOS production line to enable on-chip lasers for over 2 million optical transceiver units each year [23, 24].

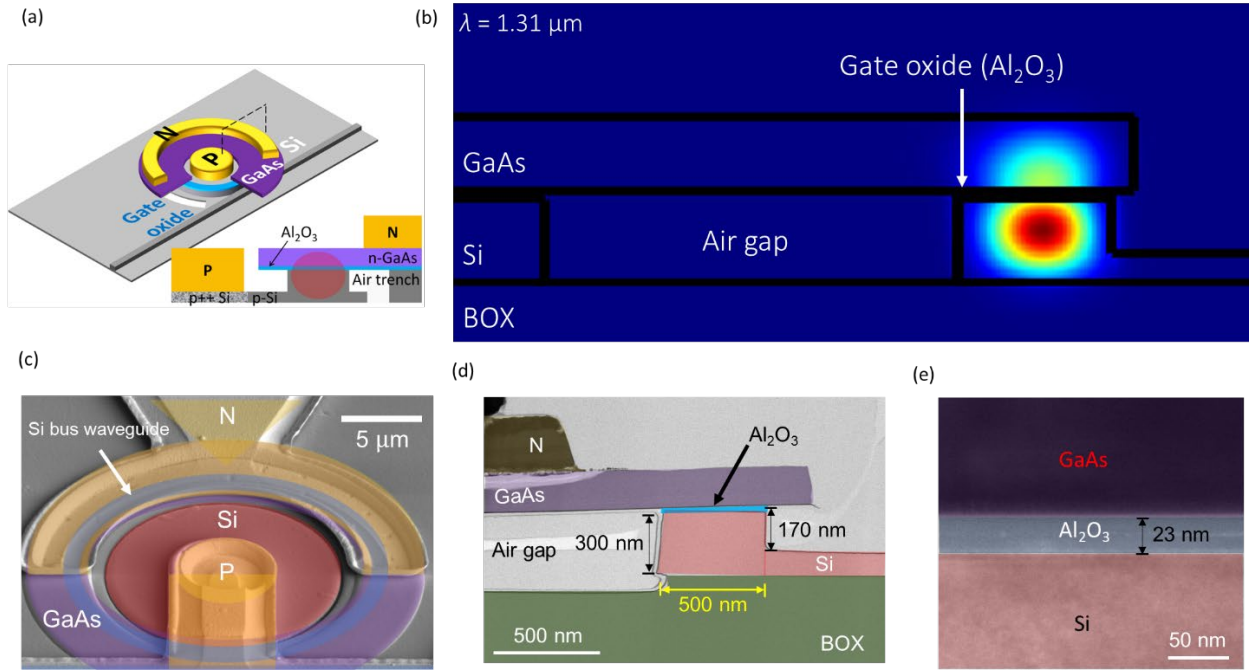

Figure 1(a) 3D-view and cross-schematic view of memristive III-V-on-Silicon microring resonator. (b) Simulated fundamental TE mode field intensity within the microring waveguide at 1310 nm. (c) Scanning electron microscopy (SEM) cross-sectional image of memresonator. (d) Transmission electron microscopy (TEM) cross-sectional image of memresonator. (e) TEM image of a bonded GaAs-Al$_2$O$_3$-Si memristor.

## Results

**Device Design and Fabrication**

As schematically shown in Figure 1(a), heterogeneous III-V/Si microring resonators (MRRs) of varying radii between 10 µm and 25 µm were fabricated on a silicon-on-insulator (SOI) substrate with a 2 µm-thick buried oxide layer and a 300 nm-thick top silicon layer. GaAs epitaxial device layers are transferred to a 100 mm Si-on-insulator (SOI) substrate by an O$_2$ plasma-assisted direct wafer bonding process [25]. About 10 nm of Al$_2$O$_3$ was grown on both the

GaAs and Si substrates using atomic layer deposition (ALD) before they were bonded together to form the resistive-switching oxide. A n-GaAs/Al$_2$O$_3$/p-Si semiconductor-insulator-semiconductor (SIS) stack is embedded within the microring resonator for high-speed optical signal modulation through carrier accumulation and the plasma dispersion effect [26]. This device can then be resistively switched like a memristor, thereby producing a memresonator, or a memristor integrated with a microring resonator, as will be discussed in further detail in the next section.

An air trench is then formed on the Si device layer with a ~170 nm waveguide rib etch depth, prior to wafer bonding in order to confine the memristor device area only to the fundamental TE mode and to minimize the area for high-speed and energy-efficient charging and discharging. The bus and ring waveguides within the microring resonators are 500 nm wide each and are separated by 200 nm at the coupling section. Figure 1(b) is an optical simulation showing the fundamental TE mode within the memresonator waveguide. Transmission electronic microscopy (TEM) images of the fully fabricated memresonator cross section and memristor material stack are shown in Figure 1(d) and (e), respectively. As seen in Fig. 1a, electrodes are placed on the 150 nm thick n-type GaAs contact layer and the 300 nm p-type Si contact layer to apply an electrical field across the oxide material. Since semiconductor materials are sandwiching the resistive-switching oxide, these memristors can be integrated within optical waveguides while adding only about 0.05 dB of insertion loss (see Supplementary Note S2), achieving much lower optical loss than with purely metal electrodes typically used in electronic memristors.

**Working Mechanism**

As mentioned in the previous section, a memristor is formed using n-type GaAs and p-type Si sandwiching a thin resistive switching $Al_2O_3$ layer. In order to resistively switch the memristor, a process creating an interchange of oxygen and semiconductor atoms, called "electroforming," must be induced by applying a high enough positive bias voltage across the memristor. The high electric field breaks some of the Al-O bonds causing oxygen atoms to migrate towards the semiconductor regions and leave behind negatively ionized vacancies within the $Al_2O_3$ layer. The oxygen vacancies form localized aluminum-rich channels, namely conductive filaments (CFs), that allow current to flow and effectively increase the conductivity of the oxide material, setting the device to a low resistance state (LRS) [27]. When a large enough electric field is applied in the opposite direction, it causes a reduction of oxygen vacancies as well as sufficient current flow to catalyze localized Joule heating, rupturing the CFs previously formed and resetting the memristor back to a high resistance state (HRS) [28, 29]. Prior studies suggest that a combination of electric field and Joule heating induces the resistive switching mechanism of $Al_2O_3$-based memristors [30]. When a positive bias (typically lower than the voltage needed for electroforming) is applied again, the CF reforms and the device switches back to a lower resistance.

      As can be seen in Figure 2(a), a schematic of the resistive switching mechanism within the memristor is shown. Oxygen vacancies are formed after electroforming, and they can be ruptured and reconnected through subsequent set and reset cycles. Figures 2(b) and (c) visually portray the carrier dynamics within the III-V/Si memristor-integrated waveguide when the memristor is in the LRS and the HRS. Figure 2(d) shows the current-voltage characteristics of the device which shows a hysteresis-type curve confirming its operation as a memristor. The

voltage was swept from 0 to 10 V and back down to 0 V, and then from 0 V to -5 V and back to 0 V to observe the hysteresis effect in the I-V characteristics. The compliance current, $I_{CC}$, was initially set to 50 µA in the forward direction and 1 mA in the reverse direction in order to prevent the device from permanent breakdown and physical damage. Typically, the device had less than 10 nA of DC leakage current in the HRS state due to high quality of the $Al_2O_3$ (Supplementary note S3). The leakage current is mostly due to trap-assisted tunneling through deep-level traps within the $Al_2O_3$ layer. The electroforming step in the memristor occurs at 9 V, the set voltage occurs at around 5 V and the reset voltage occurs around -4 V.

As can be seen in the current-voltage characteristics in Figure 2(d), the device can also be switched to an intermediate resistance state (IRS) with a resistance between the LRS and HRS by adjusting the current compliance of the measurement equipment to a value between the compliance used for the HRS and LRS. The device can also be set to multiple intermediate in this way, displaying the possibility of using these devices for analogue computing. For example, while the device is in the HRS, it can be switched to the IRS by applying a current compliance and can be switched to a LRS by applying a higher current compliance. Since a lower current compliance is applied, it physically limits the growth of the conductive filament to a certain size, thereby also limiting the device resistance. Moreover, when the memresonator is set to a low or intermediate resistance state, we found that the conduction in the memristor is observed to be diode-like, which resembles the ideal diode equation, $I \propto [e^{qV}-1]$. Since the resistive switching oxide acts as a non-degenerate semiconductor material, and each semiconductor contact layer is p- and n-doped, the device essentially acts like a p-i-n diode in which excess electrons flow from the n-type GaAs to the p-type Si and excess holes flow in the opposite direction [31]. The device begins behaving similarly to a carrier injection type modulator in which majority carriers are injected into the CF and drift from one contact region to the other through the CF. In Figure 2(c), a schematic diagram of this process is shown, displaying electrons being injected from the n-GaAs to the p-Si and holes being injected from the p-Si to the n-GaAs through the CF while the memristor is in the LRS.

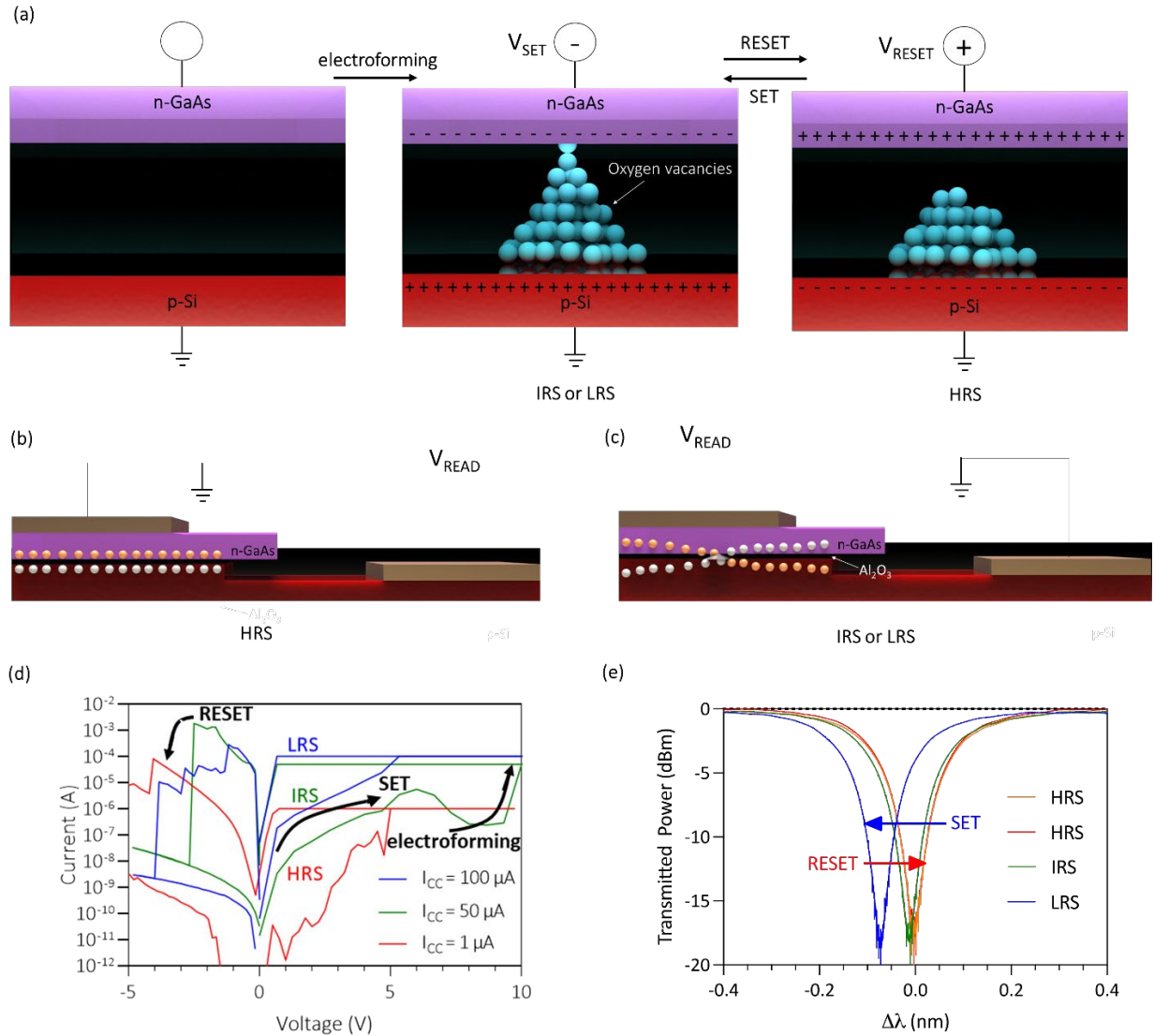

Figure 2 (a) Schematic diagram of the process of forming and rupturing conductive filaments (CFs) within the memristor. $V_{SET}$ is the voltage applied to set the memristor to the IRS or LRS. $V_{RESET}$ is the voltage applied to reset the memristor to the HRS. (b) Schematic diagram of the carrier distribution within the waveguide while a read voltage, $V_{READ}$, is applied to the memristor in the HRS. (c) Schematic diagram of the carrier distribution within the waveguide while a read voltage is applied to the memristor in the IRS or LRS. (d) Current-voltage characteristic of the device displaying the hysteresis signature of a memristor. Current compliances of 1 µA, 50 µA, and 100 µA were used in the forward bias voltage direction in order to set the device into different resistance states. (e) Optical spectrum of the memresonator while a 2 V read voltage is applied in different states.

Charge simulations were used to determine the carrier concentration as a function of the memristor state and location within the device, as well as corresponding band diagrams, can be found in Figure 3. Due to different doping levels in the GaAs and Si layers, free carriers around the Al₂O₃ oxide layer are depleted at zero bias. Without a bias voltage applied on the memristor, the doped p-Si is largely depleted because of the doped n-GaAs on the lower side of the waveguide, introducing negligible background free carrier absorption (FCA) in p-Si. Once a forward bias is applied to the memristor when in the HRS, free holes and electrons quickly accumulate in the p-Si and n-GaAs, respectively, and their concentrations increase exponentially when approaching the oxide/semiconductor interface. A schematic diagram visually portraying the carrier distribution in the HRS within the III-V/Si memristor-integrated waveguide is shown in Figure 2(b). Electrons accumulate on the interface of the Al₂O₃ layer and the n-GaAs layer, and holes accumulate at the interface of the p-Si and the Al₂O₃ layer while the memristor is the HRS.

The charge density change for electrons and holes in the HRS can be described as [32]:

$$\Delta N_e = \Delta N_h = \frac{\varepsilon_0 \varepsilon_r}{q t_{ox} t}[V - V_{FB}]$$

where $\Delta N_e$ is the change in electron concentration, $\Delta N_h$ is the change in hole concentration, $\varepsilon_0$ is the vacuum permittivity, q is the elementary charge, $\varepsilon_r$ is the relative permittivity of the oxide layer, $t_{ox}$ is the thickness of the oxide layer, t is the effective charge layer thickness, V is the applied bias voltage, and $V_{FB}$ is the flat-band voltage of the memristor in the HRS. The effective index of refraction in the waveguide decreases due to the plasma dispersion effect in the Si and GaAs layers of the waveguide and the output wavelength is reduced as a result [33]. The change in the effective index of refraction, $\Delta n_{eff}$, in the HRS can be described by the Drude-Lorentz model below [34]:

$$\Delta n_{eff} = -\frac{q^2 \lambda^2}{8\pi^2 c^2 \varepsilon_0 n}\left(\frac{\Delta N_e}{m_e} + \frac{\Delta N_h}{m_h}\right)$$

where λ is the optical wavelength, c is the velocity of light in vacuum, n is the refractive index of the material unperturbed, $m_e$ is the effective mass of electrons in the material, $m_h$ is the effective mass of holes in the material. Subsequently, this change in effective index of refraction causes a change in the resonant wavelength of the MRR [35]:

$$\Delta\lambda_r = \frac{\Delta n_{eff}\lambda_r}{n_g}$$

where $\Delta n_{eff}$ is the change in the effective index of the waveguide at the resonant wavelength, $\lambda_r$, and $n_g$ is the group index of the hybrid III-V/Si waveguide. Using a III-V material warrants a significant increase in the refractive index of the waveguide due to its high carrier mobility. By using a thinner oxide with a higher dielectric constant, such as $HfO_2$ or $TiO_2$, one could achieve a further increase in the plasma dispersion effect and tuning range [36].

Simulations show that as the carrier density within the waveguide increases, the plasma dispersion effect is enhanced and the blue shifting in the resonant wavelength occurs due to model refractive index decrease. This effect has been successfully used for energy-efficient optical phase tuning and high-speed optical modulation [26, 37]. Thermal simulations also show that with a high enough electric field, current flow through the CFs increases and catalyzes localized Joule heating (see Supplementary Note S6). This effect increases the cavity modal refractive index and counteracts the plasma dispersion effect, eventually leading to a red shift in the resonant wavelength. It is worth noting that this effect is negligible under small read voltages and that the red shifting is only observed at high read voltages (>10 V).

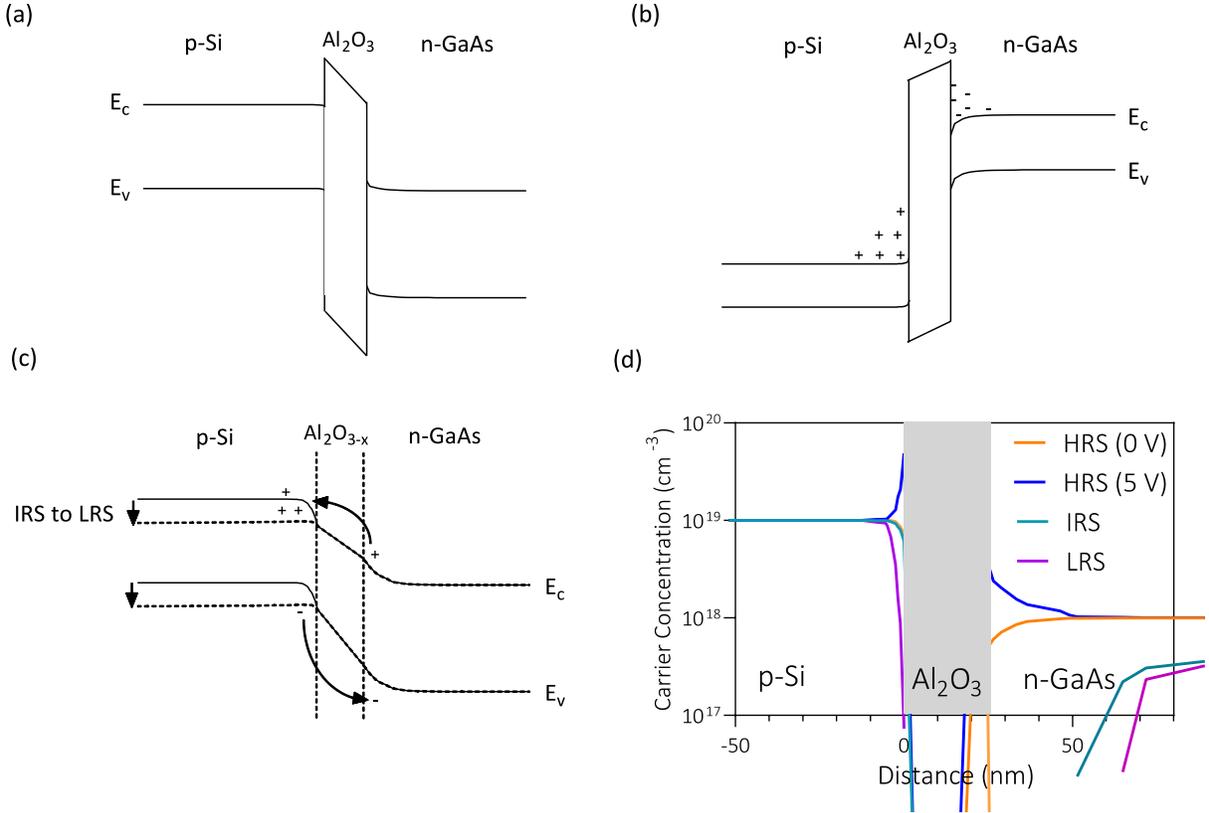

Figure 3(a) Band diagram of memristor in the HRS with zero applied bias voltage. (b) Band diagram of the memristor in the HRS with 5 V applied bias voltage. (c) Band diagram of the memristor in the LRS. (d) Simulated net carrier concentration in the memristor as a function of position in the device and bias voltage.

Furthermore, the total change in the refractive index in the IRS or LRS can be modeled using the following formula [38]:

$$\Delta \lambda_r = -\frac{\lambda_r \Gamma}{n_g}\left(n_f \Delta N - \frac{d n_g}{d T} k_{th} R \cdot I^2\right)$$

where $n_g$ is the group index of the waveguide, $n_f$ is the ratio between the change of silicon index and the change of carrier concentration when, $\Delta N$ is the change in the carrier concentration in the waveguide, $dn_g$ is the thermo-optic coefficient, $k_{th}$ is the thermal impedance of the memristor, R is the total effective series resistance, and I is the current flowing through the device.

$$\Delta N = \Delta N_p + \Delta N_i + \Delta N_n = \left(\frac{n_i^2}{N_D} + \frac{n_i^2}{N_A}\right)\left(e^{qV/kT} - 1\right) + \frac{IA\tau}{qt_{ox}}$$

The total change in the excess carrier concentration inside of the waveguide in the LRS, ΔN, can be calculated by summing the average excess carrier density inside the CF under high current level injection due to excess recombination, $\Delta N_n$, the excess electron density inside of the p-Si region, $\Delta N_p$, and the excess hole density in the n-GaAs region, $\Delta N_n$, as according to Shockley's "Law of the Junction" [39, 40]. $n_i$ is the intrinsic carrier concentration, $N_D$ is the donor concentration, q is the charge of an electron, T is the device temperature, $N_A$ is the acceptor concentration, τ is the carrier recombination lifetime inside of the CF and A is the area of the CF.

**Device Characteristics**

As shown in Figure 2(e), switching the memresonator between the LRS, IRS, and the HRS subsequently switches its resonance wavelength. The insertion loss was measured to be about 0.047 dB in the HRS and 0.048 dB in the LRS (Supplementary Note S2). The 20-μm diameter memresonator achieves about a 0.08 nm or about a 0.18π phase shift (see Supplementary Note S5) in the LRS, leading to an estimated $L_\pi$ of around 0.35 mm. The effective refractive index and phase shift as a function of voltage is plotted in Supplementary Figure S3. After setting the memresonator to the IRS or LRS, the device resonates at the same wavelength until it is reset back to the HRS. In Figure 4(a), the resistance in the HRS, IRS, and LRS and the optical power being transmitted through the memresonator at $\lambda_{HRS}$, $\lambda_{IRS}$, and $\lambda_{LRS}$ was measured for 12 hours (Figure 4(a) and (c)). This measurement demonstrates the non-volatility of this optoelectronic memory device. Drifting in the temperature stability of the setup was observed, which can be mitigated using a temperature-controlled stage. The device also demonstrated repeatability and excellent endurance as it was cycled 1,000 times between states using voltage pulses (Figure 4(b)). Figure 4(d) shows the resistance of the HRS, IRS, and LRS, and demonstrates a stable HRS/LRS resistance ratio of about $10^3$ through 1,000 switching cycles.

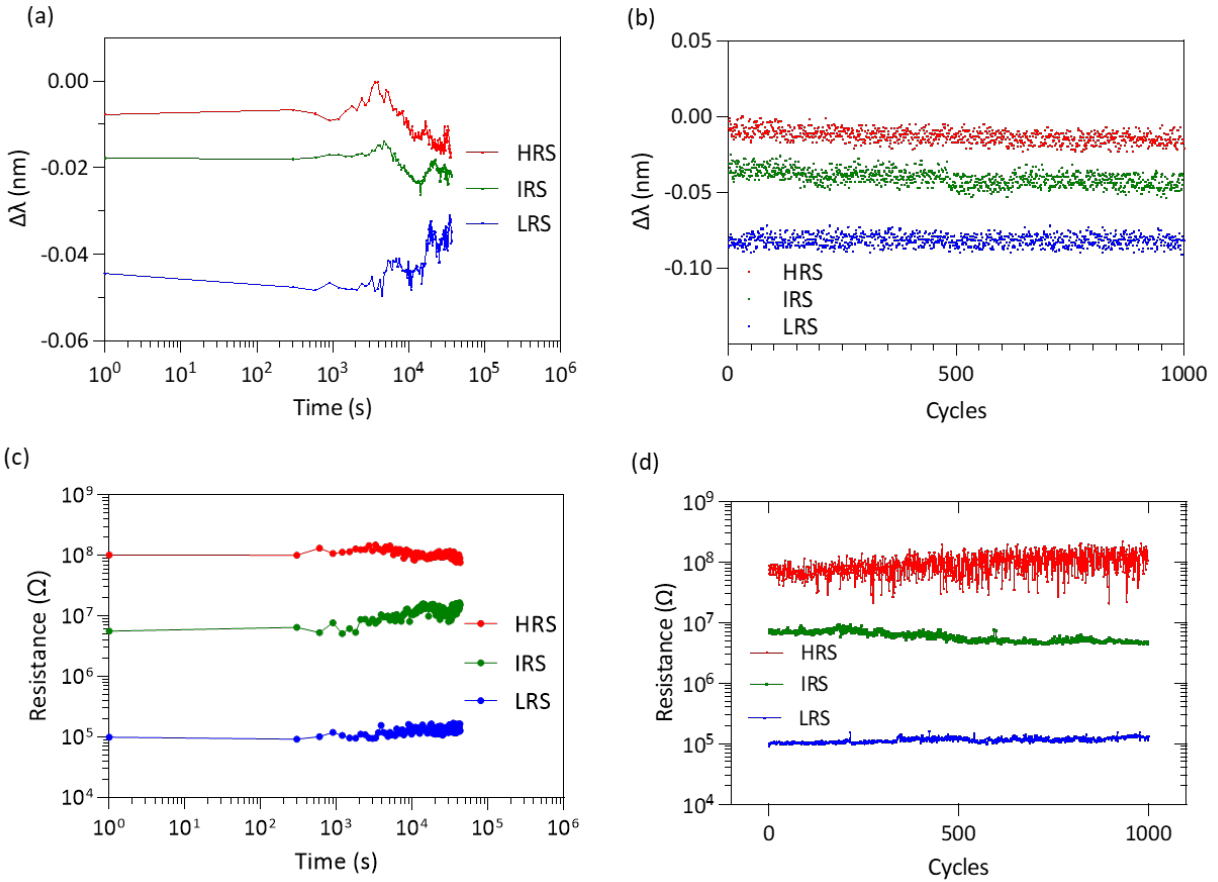

Fig. 4: (a) Wavelength shift of the memresonator in multiple states measured every 5 minutes over the span of 12 hours with a read voltage of 2 V. (b) Wavelength shift of the memresonator in multiple after 1,000 set/reset cycles with a read voltage of 3 V. (c) Resistance measurements of the memristor in multiple states monitored for 12 hours. (d) Resistance measurements of the memristor in multiple states.

To test the switching speed and energy of these devices, an arbitrary waveform generator was used to generate voltage pulses used for reading and writing the memresonator (see Methods for experiment details). The output optical power at the resonant wavelength of the memresonator was monitored as the device was being switched (Figure 5(b)-(c)). These measurements demonstrate the ability to quickly write and read data from the device with ultralow energy. The switching energy is as low as 0.15 pJ, which is more than 30x smaller than the record switching energy for photonic non-volatile memory devices [41]. After the device was SET using a 300 ps wide, 5 V amplitude pulse, then a 1 ns, 2 V read voltage was applied to read the optical power transmitted through the memresonator at the resonant wavelength as well as the read current of the memristor, which was 2.5 µA. The normalized transmitted power after the device was SET was about 0.27. The energy consumed to read the memresonator after the write cycle was about 5 fJ. Afterwards, the device was RESET using a 900 ps wide, -4 V amplitude pulse, the switching energy for RESET was measured to be 0.36 pJ.

Then a 1 ns, 2 V read voltage was applied to read the transmission power of the memresonator and the read current of the memristor, which was around 10 nA. The normalized optical power transmitted through the memresonator at the resonant wavelength after the device was SET was about 0.1. There is a small blueshift in the resonant wavelength in the HRS when the read voltage is applied, explaining why there is a small amount of power being transmitted even after the device has been RESET. However, the transmitted read power is nearly 3x times smaller than when the device has been SET. Also, the energy consumption of reading the memresonator after the erase cycle was about 2 aJ. Most importantly, zero static power is consumed in between read and write cycles as energy is only spent during the read and write operations.

The measured switching speed of these devices is over two orders of magnitude faster than the fastest non-volatile photonic phase shifters and is comparable to all-electronic memristor devices made of similar materials [42, 43]. This is consistent with literature on the switching speed of oxide-based resistive switching devices, which is typically on the timescale of hundreds of picoseconds [44]. The optoelectronic bandwidth is limited by the time required for enough Joule heating within the CF to cause it to rupture. This switching speed can be improved with optimized choice of oxide material and reducing the oxide thickness. The bandwidth of similar III-V/Si MOS-type microring modulators have been measured to be as high as 28 GHz [22, 26, 45, 46].

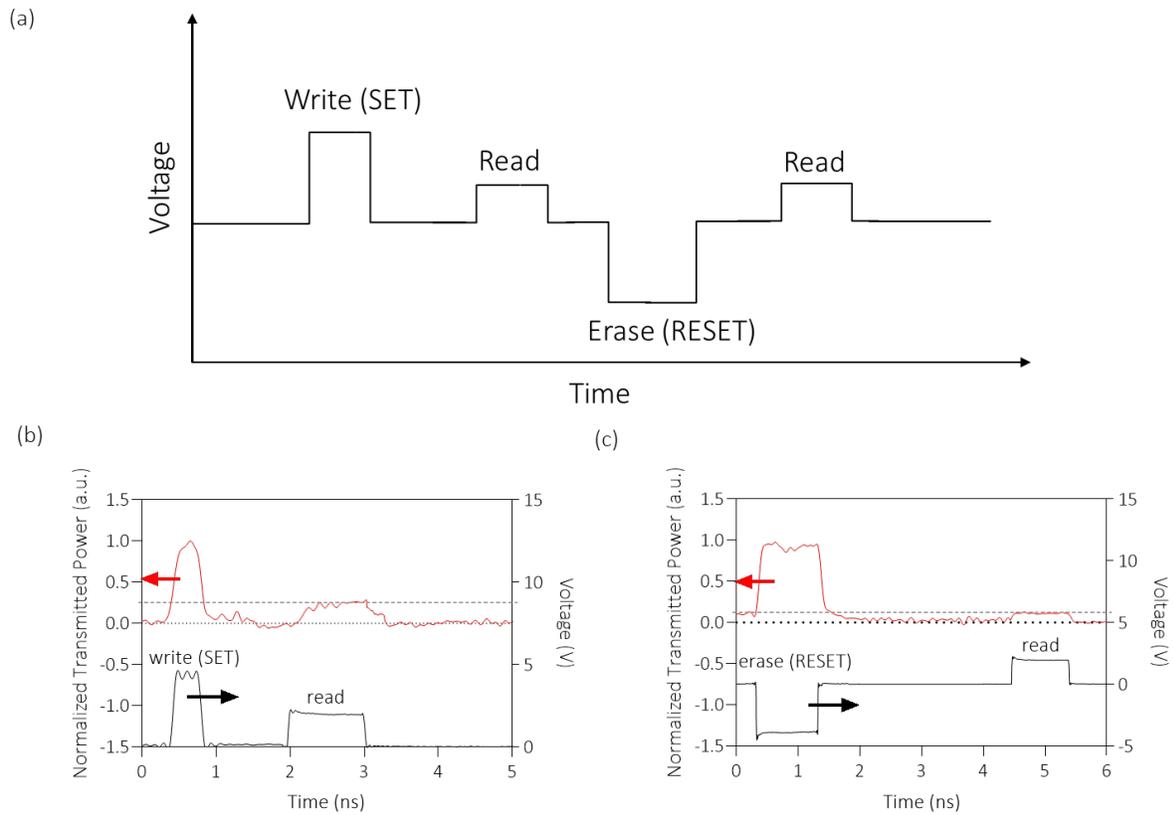

Fig. 5. (a) The typical voltage pulse sequence used to write and erase data from a memristor. (b) Plotted is the normalized transmitted power (left y-axis) at the resonant wavelength of the memresonator as a function of time during a voltage pulse sequence. The voltage of the input pulse sequence is measured on the right y-axis. The write sequence includes a 5 V amplitude, 300 ps voltage pulse used to SET the device, and a 2 V, 1 ns voltage pulse used to read. (c) The erase sequence includes a -4 V amplitude, 900 ps voltage pulse used to RESET the device, and a 2 V, 1 ns voltage pulse used to read.

**Discussion**

Resistive switching elements such as memristors have been used for analogue computing for several years. While these electronic memristors can be integrated at high densities within crossbar arrays and switched at high speeds, there exists a trade-off between bandwidth and the total size of the crossbar array. For example, the bandwidth scales inverse proportionally to the size of the crossbar array, meaning that as the size reaches greater than 1 mm$^2$, the bandwidth becomes constrained and the energy cost for off-chip communications can also become problematic. Whereas on a photonic platform, signals can be supported with much greater bandwidth and consume less energy for longer distances than the electrical counterparts. For instance, optical waveguides can be designed with low signal attenuation (<0.1dB/cm) and are able to propagate high power signals without the issues of thermal runaway such as that seen in the Joule heating of electrical wires [47]. Hence, microring-based weight banks and crossbar arrays which perform matrix-vector multiplication operations used for photonic neural networks, as well as for optical content-addressable memory, can be fashioned out of this platform using these non-volatile photonic phase shifters [8, 48, 49]. Furthermore, these types of circuits can potentially achieve larger scales with much higher efficiency than their electronic counterparts.

      Table 1 compares the characteristics for different implementations of phase shifters on silicon photonic platforms used for neural networks and optical FPGAs. Typically, thermo-optic phase shifters are used as weights within photonic neural networks, however, since they lack high-speed programming capabilities and non-volatile memory capabilities, they waste tens of mWs per unit of static power over the span of an inference task and with each weight update cycle. They can also easily cause thermal crosstalk which limits integration density and scale, and control complexity [7, 50, 51]. More recent demonstrations showed integrated nano-opto-

electro-mechanical phase shifters with improved energy efficiency, but were still limited in write speeds (~1 µs), require large switching voltages (>20 V), and have high mechanical failure rates [52, 53, 54].

On the other hand, phase-change materials (PCM) such as $Ge_2Sb_2Te_5$ (GST) and, more recently, $Sb_2Se_3$ have been explored extensively as a candidate for non-volatile memory within silicon PICs with encouraging results [41, 55, 56, 57, 58]. However, these materials are also limited in writing speed and typically require high input powers (~mW) to heat them long enough to change the phase from amorphous to crystalline. Most recently, $BaTiO_3$ (BTO) non-volatile phase shifters have also been demonstrated with multi-level states with a switching energy as low as 4.6 pJ and excellent controllability [42]. While a promising advancement, they require a reset sequence consisting of 10,000 pulses with a duration totaling hundreds of microseconds before switching states. Phase shifters based on BTO also require about a ~1 mm phase shifter length to achieve a π phase shift, which is challenging to scale and achieve high speed operation. It is worth nothing that the BTO phase shifters were implemented on Mach-Zehnder Interferometers, which inherently have a larger device footprint than microring resonators.

|  | Thermo-optic [50, 51, 59] | Charge-trapping [60, 61] | Plasmonics [62, 63] | MEMS [52, 64] | PCM [10, 41, 56, 57, 65, 66] | BaTiO$_3$ (BTO) [42, 67] | Memresonator (This work) |
|---|---|---|---|---|---|---|---|
| **Switching Speed** | 5-200 $\mu$s | > 350 ms | ~10 $\mu$s | ~1 $\mu$s | < 100 ns | < 1 ms | < 1 ns |
| **Switching Energy** | 1$\mu$J | 11.4-17.2 pJ | 1 nJ | 0.2 nJ | 13.4 pJ | 4.6-26.7 pJ | 0.15-0.36 pJ |
| **Retention Time** | N/A | 10 years | – | N/A | 10 years | 10 hours | 12 hours |
| **Device Footprint** | ~100 $\mu$m | 20 $\mu$m | 5 $\mu$m | 30 $\mu$m | 33 $\mu$m | 150 $\mu$m | 20 $\mu$m |
| **Insertion Loss** | 6.5 dB | ~1 dB | ~4 dB | 3.5 dB | ~1 dB | ~0.1 dB | 0.048 dB |
| **Non-volatility** | No | Yes | Yes | No | Yes | Yes | Yes |

Table 1: Implementations of programmable phase shifters on a silicon photonic platform.

In this work, we have demonstrated a non-volatile III-V-on-silicon memresonator used for programmable photonic memory operating at record low switching energy (0.15-0.36 pJ), sub-nanosecond switching times enabling high-speed, energy-efficient in-memory computing within silicon photonic neural networks. These non-volatile optoelectronic memory devices save a great deal of energy by reducing the power consumption involved in programming phase shifters within photonic integrated circuits. By using short voltage pulses to permanently switch the state of this device, the energy typically lost in idle power consumption is saved throughout the duration of an inference task. For example, after a write pulse is applied to the memresonator, the device will retain its state until another voltage pulse is used to write a different weight value. In this way, it is worth reiterating that no idle power consumption is wasted in between reading or writing the weight value stored within the memresonator.

Additionally, these non-volatile photonic phase shifters can act as weights within silicon photonic neural networks which can be updated in real-time, enabling algorithms such as error back-propagation to be executed directly on-chip, greatly optimizing the acceleration of silicon photonic neural networks. Another significant distinction is that the memory is directly on the same chip as the phase shifter, enabling the capability to perform in-memory photonic computing. This avoids the optical to electrical conversion losses involved in going to off-chip memory in between each data set used to train a neural network. For instance, given that training typically requires frequently retrieving results from intermediate layers, this would save a substantial amount of energy involved in fetching those results typically stored in an external memory chip like static random-access memory (SRAM) or dynamic random-access memory (DRAM). Another specific type of neural network this may apply to is transfer learning, described as the practice of re-using a pre-trained neural network instead of training one from scratch to reduce latency and save computational resources. Given that the weights in the backbone layer are fixed, they would benefit from being stored in on-chip memory such as with these memresonators. Also, these memresonators can simultaneously be used for the trainable portion of the neural network since they are also capable of being updated at high speeds and energy-efficiency.

Lastly, these devices were developed on a heterogeneous III-V-on-silicon platform, which allows for the co-integration of non-linear active optoelectronic devices, such as lasers and modulators, directly on the same chip as a silicon photonic neural network or an optical FPGA [68]. Since these types of photonic integrated circuits do not inherently need to transmit optical signals off-chip, we gain a significant advantage by integrating the light source directly on-chip. This technology can immensely improve the energy-efficiency, stability, and scalability of integrated photonic processors, advancing their potential for use in next-generation HPCs and edge computing.

Future designs will feature device and structural design improvements in order to reduce switching voltages. The total voltage applied across the device distributes over the $Al_2O_3$ layer, semiconductor layers (n-GaAs, p-Si), and metal/semiconductor contact layers. By reducing the active area of the device and thickness of the $Al_2O_3$ layer, optimizing the semiconductor layers' doping concentrations and thicknesses, and improving the quality of the metal/semiconductor contact interface and the $Al_2O_3$ layer, we can reduce the switching voltage [69]. Another design change will be to integrate a field-effect transistor in series with the memristor to be able to apply voltage pulses on the device with control of device current. Also, TEM images will be taken to investigate the conductive filament formation within these devices and study the physical processes behind the switching mechanisms in these devices. These studies will aid in the design of future devices such as the selection of the resistive-switching oxide material.

Additionally, these devices can be also integrated with a MOS field-effect transistor (MOSFET) in a one-transistor one-resistor (1T1R) configuration to reliably control the current flow in the device without external circuitry. Within a 1T1R configuration, a MOSFET is connected in series with a memristor and is used to limit the current in the memristor by applying a gate voltage on the MOSFET to modulate the channel length and allow only a certain amount of current to flow through the MOSFET channel. Lastly, memristors can also be integrated within Mach-Zehnder interferometers as an alternative form of a non-volatile phase shifter also commonly used within silicon photonic neural networks, quantum computing circuits, and FPGAs [70].

## Methods

Microring resonators with a diameter of 20 μm were measured on a copper stage with III-V side up. The experimental set ups for the measurements taken are shown in Fig. 6. Electrical measurements were taken with an Agilent B1500A semiconductor device analyzer including a B1525A HV-SPGU high-voltage pulse generator. GSG RF probes (CascadeMicrotech ACP-40) were used to probe the devices and measure the high-speed response. Optical power measurements were taken using a Newport 2936-R optical power meter. The device was designed with input and output grating couplers, which had about 6 dB of loss each at peak transmission. A Santec TSL-510 tunable laser is used to illuminate the input grating coupler with a cleaved fiber. The laser wavelength is swept and the output of the device is measured through the output grating coupler which is coupled to an optical power meter.

    In order to measure the switching speed of the memresonator, we couple light coming from a tunable laser at the resonant wavelength of the memresonator into the input grating coupler. We then apply voltage pulses from a Keysight M8195A Arbitrary Waveform Generator to read and write the memristor, and couple light coming from the output grating coupler into a high-speed photodiode which is then connected to a Tektronix 8 GHz real-time oscilloscope. A 100 ns wide, 2 V amplitude pulse was used to read the memresonator in the retention time and endurance measurements.

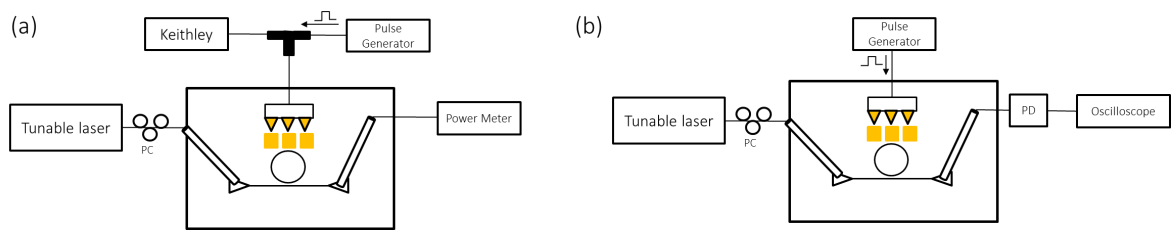

Figure 6. (a) Schematic diagram of the experimental setup used during the endurance and retention time measurements. (b) Schematic diagram of the experimental setup used during the of the switching speed measurement.

# Acknowledgments

The authors want to thank Thomas Van Vaerenbergh, Marco Fiorentino, and Sri Priya Sundararajan for the insightful discussions and input towards this paper.

# Author contributions

Conceptualization: BT, DL, JPS

Methodology: BT, DL, SC, ZF, XS

Investigation: BT, DL, SC, ZF, XS

Visualization: BT, ZF

Supervision: DL, RGB

Writing—original draft: BT

Writing—review & editing: BT, DL, ZF, RGB

**Competing interests:** All other authors declare they have no competing interests.

**Data and materials availability:** All data are available in the main text or the supplementary materials.

**Funding statement:** The authors acknowledge that they received no funding in support for this research.

**References**

[1] D. e. a. Amodei, *AI and Compute,* 2018.
[2] *IBM's New Brain (IEEE Spectrum) – Jeremy Hsu.*
[3] *See https://mythic.ai/technology/ for Mythic..*
[4] B. J. Shastri, A. N. Tait, T. Ferreira de Lima, W. H. P. Pernice, H. Bhaskaran, C. D. Wright and P. R. Prucnal, "Photonics for artificial intelligence and neuromorphic computing," *Nature Photonics,* vol. 15, p. 102–114, February 2021.
[5] M. A. Taubenblatt, "Optical interconnects for high performance computing," in *IEEE Photonic Society 24th Annual Meeting*, 2011.
[6] F. Ashtiani, A. J. Geers and F. Aflatouni, "An on-chip photonic deep neural network for image classification," *Nature,* vol. 606, p. 501–506, June 2022.
[7] Y. Shen, N. C. Harris, S. Skirlo, M. Prabhu, T. Baehr-Jones, M. Hochberg, X. Sun, S. Zhao, H. Larochelle, D. Englund and M. Soljačić, "Deep learning with coherent nanophotonic circuits," *Nature Photonics,* vol. 11, p. 441–446, July 2017.
[8] A. N. Tait, T. F. de Lima, E. Zhou, A. X. Wu, M. A. Nahmias, B. J. Shastri and P. R. Prucnal, "Neuromorphic photonic networks using silicon photonic weight banks," *Scientific Reports,* vol. 7, p. 7430, August 2017.
[9] B. Shi, N. Calabretta and R. Stabile, "InP photonic integrated multi-layer neural networks: Architecture and performance analysis," *APL Photonics,* vol. 7, p. 010801, January 2022.
[10] J. Feldmann, N. Youngblood, C. D. Wright, H. Bhaskaran and W. H. P. Pernice, "All-optical spiking neurosynaptic networks with self-learning capabilities," *Nature,* vol. 569, p. 208–214, May 2019.
[11] E. A. Vlieg, L. Talandier, R. Dangel, F. Horst and B. J. Offrein, "An Integrated Photorefractive Analog Matrix-Vector Multiplier for Machine Learning," *Applied Sciences,* vol. 12, p. 4226, January 2022.
[12] M. A. Nahmias, A. N. Tait, B. J. Shastri, T. F. de Lima and P. R. Prucnal, "Excitable laser processing network node in hybrid silicon: analysis and simulation," *Optics Express,* vol. 23, p. 26800, October 2015.
[13] S. Pai, B. Bartlett, O. Solgaard and D. A. B. Miller, "Matrix Optimization on Universal Unitary Photonic Devices," *Physical Review Applied,* vol. 11, p. 064044, June 2019.
[14] T. W. Hughes, M. Minkov, Y. Shi and S. Fan, "Training of photonic neural networks through in situ backpropagation and gradient measurement," *Optica,* vol. 5, p. 864–871, July 2018.
[15] L. Chua, "Memristor-The missing circuit element," *IEEE Transactions on Circuit Theory,* vol. 18, p. 507–519, September 1971.
[16] D. B. Strukov, G. S. Snider, D. R. Stewart and R. S. Williams, "The missing memristor found," *Nature,* vol. 453, p. 80–83, May 2008.

[17] A. C. Torrezan, J. P. Strachan, G. Medeiros-Ribeiro and R. S. Williams, "Sub-nanosecond switching of a tantalum oxide memristor," *Nanotechnology,* vol. 22, p. 485203, November 2011.

[18] J. J. Yang, M.-X. Zhang, J. P. Strachan, F. Miao, M. D. Pickett, R. D. Kelley, G. Medeiros-Ribeiro and R. S. Williams, "High switching endurance in TaOx memristive devices," *Applied Physics Letters,* vol. 97, p. 232102, December 2010.

[19] M.-J. Lee, C. B. Lee, D. Lee, S. R. Lee, M. Chang, J. H. Hur, Y.-B. Kim, C.-J. Kim, D. H. Seo, S. Seo, U.-I. Chung, I.-K. Yoo and K. Kim, "A fast, high-endurance and scalable non-volatile memory device made from asymmetric Ta2O5−x/TaO2−x bilayer structures," *Nature Materials,* vol. 10, p. 625–630, August 2011.

[20] B. Govoreanu, G. S. Kar, Y.-Y. Chen, V. Paraschiv, S. Kubicek, A. Fantini, I. P. Radu, L. Goux, S. Clima, R. Degraeve, N. Jossart, O. Richard, T. Vandeweyer, K. Seo, P. Hendrickx, G. Pourtois, H. Bender, L. Altimime, D. J. Wouters, J. A. Kittl and M. Jurczak, "10x10nm2 Hf/HfOx Crossbar Resistive RAM with Excellent Performance, Reliability and Low-Energy Operation," p. 4.

[21] D. B. Strukov and R. S. Williams, "Four-dimensional address topology for circuits with stacked multilayer crossbar arrays," *Proceedings of the National Academy of Sciences,* vol. 106, p. 20155–20158, December 2009.

[22] D. Liang, S. Srinivasan, G. Kurczveil, B. Tossoun, S. Cheung, Y. Yuan, A. Descos, Y. Hu, Z. Huang, P. Sun, T. Van Vaerenbergh, C. Zhang, X. Zeng, S. Liu, J. E. Bowers, M. Fiorentino and R. G. Beausoleil, "An Energy-Efficient and Bandwidth-Scalable DWDM Heterogeneous Silicon Photonics Integration Platform," *IEEE Journal of Selected Topics in Quantum Electronics,* vol. 28, p. 1–19, November 2022.

[23] R. Jones, P. Doussiere, J. B. Driscoll, W. Lin, H. Yu, Y. Akulova, T. Komljenovic and J. E. Bowers, "Heterogeneously Integrated InP\textbackslash/Silicon Photonics: Fabricating Fully Functional Transceivers," *IEEE Nanotechnology Magazine,* vol. 13, p. 17–26, April 2019.

[24] N. Margalit, C. Xiang, S. M. Bowers, A. Bjorlin, R. Blum and J. E. Bowers, "Perspective on the future of silicon photonics and electronics," *Applied Physics Letters,* vol. 118, p. 220501, May 2021.

[25] D. Liang and J. E. Bowers, "Highly efficient vertical outgassing channels for low-temperature InP-to-silicon direct wafer bonding on the silicon-on-insulator substrate," *J. Vac. Sci. Technol. B,* vol. 26, p. 9, 2008.

[26] S. Srinivasan, D. Liang and R. G. Beausoleil, "Heterogeneous SISCAP Microring Modulator for High-Speed Optical Communication," in *2020 European Conference on Optical Communications (ECOC)*, 2020.

[27] R. Waser, R. Dittmann, G. Staikov and K. Szot, "Redox-Based Resistive Switching Memories – Nanoionic Mechanisms, Prospects, and Challenges," *Advanced Materials,* vol. 21, p. 2632–2663, 2009.

[28] W. Sun, "Understanding memristive switching via in situ characterization and device modeling," *Nature Communications,* vol. 10, p. 13, 2019.

[29] Y. Zhang, "Evolution of the conductive filament system in HfO2-based memristors observed by direct atomic-scale imaging," *Nature Communications,* vol. 12, p. 10, 2021.

[30] X. Zhang, L. Xu, H. Zhang, J. Liu, D. Tan, L. Chen, Z. Ma and W. Li, "Effect of Joule Heating on Resistive Switching Characteristic in AlOx Cells Made by Thermal Oxidation Formation," *Nanoscale Research Letters,* vol. 15, p. 11, January 2020.

[31] C. Li, L. Han, H. Jiang, M.-H. Jang, P. Lin, Q. Wu, M. Barnell, J. J. Yang, H. L. Xin and Q. Xia, "Three-dimensional crossbar arrays of self-rectifying Si/SiO2/Si memristors," *Nature Communications,* p. 9.

[32] A. Liu, R. Jones, L. Liao, D. Samara-Rubio, D. Rubin, O. Cohen, R. Nicolaescu and M. Paniccia, "A high-speed silicon optical modulator based on a metal–oxide– semiconductor capacitor," vol. 427, p. 4, 2004.

[33] D. Liang, X. Huang, G. Kurczveil, M. Fiorentino and R. G. Beausoleil, "Integrated finely tunable microring laser on silicon," *Nature Photonics,* vol. 10, p. 719–722, November 2016.

[34] R. Soref and B. Bennett, "Electrooptical effects in silicon," *IEEE Journal of Quantum Electronics,* vol. 23, p. 123–129, January 1987.

[35] W. Bogaerts, P. D. Heyn, T. Van Vaerenbergh, K. D. Vos, S. Kumar, T. Claes, P. Dumon, P. Bienstman, D. V. Thourhout and R. Baets, "Silicon microring resonators," p. 28, 2011.

[36] X. Huang, D. Liang, C. Zhang, G. Kurczveil, X. Li, J. Zhang, M. Fiorentino and R. Beausoleil, "Heterogeneous MOS microring resonators," in *2017 IEEE Photonics Conference (IPC)*, 2017.

[37] T. Hiraki, T. Aihara, K. Hasebe, K. Takeda, T. Fujii, T. Kakitsuka, T. Tsuchizawa, H. Fukuda and S. Matsuo, "Heterogeneously integrated III–V/Si MOS capacitor Mach–Zehnder modulator," *Nature Photonics,* vol. 11, p. 482–485, August 2017.

[38] R. Wu, C.-H. Chen, J.-M. Fedeli, M. Fournier, K.-T. Cheng and R. G. Beausoleil, "Compact models for carrier-injection silicon microring modulators," p. 10, 2015.

[39] A. Herlet, "The forward characteristic of silicon power rectifiers at high current densities," *Solid-State Electronics,* vol. 11, p. 717–742, August 1968.

[40] W. Shockley, "The theory of p-n junctions in semiconductors and p-n junction transistors," *The Bell System Technical Journal,* vol. 28, p. 435–489, July 1949.

[41] C. Ríos, M. Stegmaier, P. Hosseini, D. Wang, T. Scherer, C. D. Wright, H. Bhaskaran and W. H. P. Pernice, "Integrated all-photonic non-volatile multi-level memory," *Nature Photonics,* vol. 9, p. 725–732, November 2015.

[42] J. Geler-Kremer, F. Eltes, P. Stark, D. Stark, D. Caimi, H. Siegwart, B. Jan Offrein, J. Fompeyrine and S. Abel, "A ferroelectric multilevel non-volatile photonic phase shifter," *Nature Photonics,* vol. 16, p. 491–497, July 2022.

[43] H. Y. Lee, P. S. Chen, T. Y. Wu, Y. S. Chen, C. C. Wang, P. J. Tzeng, C. H. Lin, F. Chen, C. H. Lien and M.-J. Tsai, "Low power and high speed bipolar switching with a thin reactive Ti buffer layer in robust HfO2 based RRAM," in *2008 IEEE International Electron Devices Meeting*, 2008.

[44] S. Menzel, M. v. Witzleben, V. Havel and U. Böttger, "The ultimate switching speed limit of redox-based resistive switching devices," *Faraday Discussions,* vol. 213, p. 197–213, February 2019.

[45] D. Liang, C. Zhang, A. Roshan-Zamir, K. Yu, C. Li, G. Kurczveil, Y. Hu, W. Shen, M. Fiorentino, S. Kumar, S. Palermo and R. Beausoleil, "A Fully-integrated Multi-λ Hybrid

DML Transmitter," in *2018 Optical Fiber Communications Conference and Exposition (OFC)*, 2018.

[46] T. Thiessen, P. Grosse, J. D. Fonseca, P. Billondeau, B. Szelag, C. Jany, J. k. S. Poon and S. Menezo, "30 GHz heterogeneously integrated capacitive InP-on-Si Mach-Zehnder modulators," *Optics Express,* vol. 27, p. 102–109, January 2019.

[47] T. F. d. L. A. N. T. H.-T. P. B. J. S. P. R. P. Mitchell A. Nahmias, "Photonic Multiply-Accumulate Operations for Neural Networks," *IEEE Journal of Selected Topics in Quantum Electronics,* vol. 26, no. 1, 2020.

[48] X. Xiao, S. Cheung, S. Hooten, Y. Peng, B. Tossoun, T. Van Vaerenbergh, G. Kurczveil, R. Beausoleil, "Wavelength-Parallel Photonic Tensor Core Based on Multi-FSR Microring Resonator Crossbar Array," in *Optical Fiber Communication Conference (OFC) 2023*, San Diego, 2023.

[49] Y. London, T. Van Vaerenbergh, L. Ramini, C. Li, C. E. Graves, M. Fiorentino and R. G. Beausoleil, "WDM Ternary Content-Addressable Memory For Optical Links," in *2023 IEEE 18th International Conference on Group IV Photonics (GFP)*, Arlington, Virginia, 2023.

[50] M. J. R. Heck, "Highly integrated optical phased arrays: photonic integrated circuits for optical beam shaping and beam steering," *Nanophotonics,* vol. 6, p. 93–107, January 2017.

[51] C. Zhong, H. Ma, C. Sun, M. Wei, Y. Ye, B. Tang, P. Zhang, R. Liu, J. Li, L. Li and H. Lin, "Fast thermo-optical modulators with doped-silicon heaters operating at 2 μm," *Optics Express,* vol. 29, p. 23508–23516, July 2021.

[52] M. Dong, G. Clark, A. J. Leenheer, M. Zimmermann, D. Dominguez, A. J. Menssen, D. Heim, G. Gilbert, D. Englund and M. Eichenfield, "High-speed programmable photonic circuits in a cryogenically compatible, visible–near-infrared 200 mm CMOS architecture," *Nature Photonics,* vol. 16, p. 59–65, January 2022.

[53] R. Baghdadi, M. Gould, S. Gupta, M. Tymchenko, D. Bunandar, C. Ramey and N. C. Harris, "Dual slot-mode NOEM phase shifter," *Optics Express,* vol. 29, p. 19113–19119, June 2021.

[54] C. Lee, "Reliability and failure analysis of MEMS/NEMS switches," in *2016 IEEE 23rd International Symposium on the Physical and Failure Analysis of Integrated Circuits (IPFA)*, 2016.

[55] Z. Fang, R. Chen, J. Zheng and A. Majumdar, "Non-Volatile Reconfigurable Silicon Photonics Based on Phase-Change Materials," *IEEE Journal of Selected Topics in Quantum Electronics,* vol. 28, p. 1–17, October 2021.

[56] M. Wuttig, H. Bhaskaran and T. Taubner, "Phase-change materials for non-volatile photonic applications," *Nature Photonics,* vol. 11, p. 465–476, August 2017.

[57] J. Wang, L. Wang and J. Liu, "Overview of Phase-Change Materials Based Photonic Devices," *IEEE Access,* vol. 8, p. 121211–121245, 2020.

[58] Z. Fang, R. Chen, J. Zheng, A. I. Khan, K. M. Neilson, S. J. Geiger, D. M. Callahan, M. G. Moebius, A. Saxena, M. E. Chen, C. Rios, J. Hu, E. Pop and A. Majumdar, "Ultra-low-energy programmable non-volatile silicon photonics based on phase-change materials with graphene heaters," *Nature Nanotechnology,* p. 1–7, July 2022.

[59] S. Liu, J. Feng, Y. Tian, H. Zhao, L. Jin, B. Ouyang, J. Zhu and J. Guo, "Thermo-optic phase shifters based on silicon-on-insulator platform: state-of-the-art and a review," *Frontiers of Optoelectronics,* vol. 15, p. 9, 2022.

[60] I. Olivares, J. Parra and P. Sanchis, "Non-Volatile Photonic Memory Based on a SAHAS Configuration," *IEEE Photonics Journal,* vol. 13, p. 1–8, April 2021.

[61] J.-F. Song, X.-S. Luo, A. E.-J. Lim, C. Li, Q. Fang, T.-Y. Liow, L.-X. Jia, X.-G. Tu, Y. Huang, H.-F. Zhou and G.-Q. Lo, "Integrated photonics with programmable non-volatile memory," *Scientific Reports,* vol. 6, p. 22616, March 2016.

[62] C. Hoessbacher, Y. Fedoryshyn, A. Emboras, A. Melikyan, M. Kohl, D. Hillerkuss, C. Hafner and J. Leuthold, "The plasmonic memristor: a latching optical switch," *Optica,* vol. 1, p. 198–202, October 2014.

[63] A. Emboras, I. Goykhman, B. Desiatov, N. Mazurski, L. Stern, J. Shappir and U. Levy, "Nanoscale Plasmonic Memristor with Optical Readout Functionality," *Nano Letters,* vol. 13, p. 6151–6155, December 2013.

[64] N. Quack, H. Sattari, A. Y. Takabayashi, Y. Zhang, P. Verheyen, W. Bogaerts, P. Edinger, C. Errando-Herranz and K. B. Gylfason, "MEMS-Enabled Silicon Photonic Integrated Devices and Circuits," *IEEE Journal of Quantum Electronics,* vol. 56, p. 1–10, February 2020.

[65] Z. Cheng, C. Ríos, W. H. P. Pernice, C. D. Wright and H. Bhaskaran, "On-chip photonic synapse," *Science Advances,* vol. 3, p. e1700160, September 2017.

[66] M. Stegmaier, C. Ríos, H. Bhaskaran, C. D. Wright and W. H. P. Pernice, "Nonvolatile All-Optical 1 × 2 Switch for Chipscale Photonic Networks," *Advanced Optical Materials,* vol. 5, p. 1600346, 2017.

[67] J. Capmany and D. Pérez-López, "A new change of phase," *Nature Photonics,* vol. 16, p. 479–480, July 2022.

[68] X. Xiao, M. B. On, T. Van Vaerenbergh, D. Liang, R. G. Beausoleil and S. J. B. Yoo, "Large-scale and energy-efficient tensorized optical neural networks on III–V-on-silicon MOSCAP platform," *APL Photonics,* p. 12, 2021.

[69] B. Tossoun, X. Sheng, J. Paul Strachan, D. Liang and R. G. Beausoleil, "Hybrid silicon MOS optoelectronic memristor with non-volatile memory," in *2020 IEEE Photonics Conference (IPC)*, 2020.

[70] S. Cheung, B. Tossoun, Y. Yuan, Y. Peng, G. Kurczveil, Y. Hu, X. Xian, D. Liang, and R. G. Beausoleil, "Heterogeneous III-V/Si Non-Volatile Optical Memory: A Mach-Zehnder Memristor," in *Conference on Lasers and Electro-Optics*, San Jose, 2022.

# Supplementary Materials for

## High-Speed and Energy-Efficient Non-Volatile Silicon Photonic Memory Based on Heterogeneously Integrated Memresonator

Bassem Tossoun *et al.*

*Corresponding author. Email: bassem.tossoun@hpe.com

**This PDF file includes:**

Supplementary Text
Figs. S1 to S6
Supplementary Text References

**Supplementary Text**

S1. Switching Energy

The switching energy can then be calculated by integrating the product of $I_{memristor}$ x $V_{memristor}$ under the duration of the voltage pulse. In the case of the SET cycle, $V_{memristor} = V_{SET}$, the amplitude of the voltage pulse used to SET the memristor into the LRS, and $I_{memristor} = I_{SET}$, the maximum current in the memristor when the device switches to the LRS. In the case of the RESET cycle, $V_{memristor} = V_{RESET}$, the amplitude of the voltage pulse used to RESET the memristor into the LRS, and $I_{memristor} = I_{RESET}$, the peak absolute current before the current decreases due to switching. The current values are measured with a Keysight B1500A source measuring unit while pulses are sent to the device with the Keysight B1525A pulse generator unit.

To take account of the variation of switching conditions across devices, we averaged the switching power across 4 devices, all having been switched using 4 V and 5 V, 100 ns voltage pulses and 5 ns rise and fall times. These values can also be extracted from the resistance measurements shown in Figure 4(d) in the manuscript. The average switching power for SET is the product of $I_{memristor}$ x $V_{memristor}$ = 100 µA x 5 V = 500 µW ± 80 µW and the average switching energy for SET is 500 µW x 300 ps = 0.15 pJ ± 0.3 pJ. The average switching power for RESET is 100 µA x 4 V = 400 µW ± 90 µW and the average switching energy for RESET is 400 µW x 900 ps = 0.36 pJ ± 0.08 pJ. In order to read the device, it costs less than 2.5 µA x 2 V = 5 µW of power and only 5 µW x 1 ns = 5 fJ of energy when reading in the LRS, and 1 nA x 2 V = 2 nW and only 2 nW x 1 ns = 2 aJ.of energy to read in the HRS.

When used in practice, the memristor typically operates using voltage pulse sequences: a write followed by a read cycle, and an erase followed by a read cycle. One voltage pulse is used to switch the state and write/erase data onto the memristor, and another pulse is used to read the current in the memristor. In the case of the memristor, a read voltage is applied to read the resonant wavelength of the device. In between read and write cycles, there is zero static power consumption.

## S2. Insertion and Coupling Loss

In order to calculate the coupling and insertion losses within the memresonator, we used the following equations:

$$\mathcal{T} \equiv \left(\frac{t^2 + \alpha^2 - 2\alpha t \cos\phi}{1 + \alpha^2 t^2 - 2\alpha t \cos\phi}\right)$$

$$\mathcal{F} \equiv \frac{\Delta\lambda_{FSR}}{\Delta\lambda_{FWHM}}$$

$$\mathcal{E} \equiv \frac{T_{max}}{T_{min}}$$

$$\mathcal{E} = \left[\frac{(\alpha+t)(1-\alpha t)}{(\alpha-t)(1+\alpha t)}\right]^2$$

$$\cos(\pi/\mathcal{F}) = \frac{2\alpha t}{1 + \alpha^2 t^2}$$

$$A \equiv \frac{\cos(\pi/\mathcal{F})}{1 + \sin(\pi/\mathcal{F})}$$

$$B \equiv 1 - \left[\frac{1-\cos(\pi/\mathcal{F})}{1+\cos(\pi/\mathcal{F})}\right]\frac{1}{\mathcal{E}}$$

$$(\alpha, t) = \left(\frac{A}{B}\right)^{1/2} \pm \left(\frac{A}{B} - A\right)^{1/2}$$

$\mathcal{F}$ is the finesse, $\mathcal{E}$ is the extinction ratio, $\alpha$ is the loss coefficient, or the optical loss in one roundtrip of the microring resonator, t is the coupling coefficient, FSR is the free spectral range of the microring resonator, and FWHM is the full-width at half maximum of the resonance. In our case the FSR was around 2.808 nm and the FWHM was around 0.13 nm, giving a finesse of around 21.5. The Q-factor of the ring is around 9,933 and the extinction ratio is 27.55.

A Lorentzian fit was used to curve fit the resonances for extraction of the coefficients and the resonance wavelength throughout measurements made. The function used is shown below:

$$f(\lambda) = A_{res}\left(1 - \left(1 - \frac{1}{rE}\right)\frac{(1/2(\text{FWHM}))^2}{(\lambda - \lambda_0)^2 + (1/2(\text{FWHM}))^2}\right)$$

where $A_{res}$ is the normalization factor of the resonance, rE is the extinction coefficient, FWHM is the full width at half maximum, and $\lambda_0$ is the mean wavelength of the Lorentzian fit. Fig. S3 displays a plot of a resonance including the Lorentzian fit.

Fig. S4 are plots including the different memresonator structures that we measured and their corresponding coupling and insertion losses. With $\alpha = 0.945$, we can calculate the insertion loss in the HRS while the device is being read through the follow equation:

$IL = 10 * log_{10} \alpha = 0.047$ dB

In the LRS, α = 0.937, we can calculate the insertion loss to be while the device is being read:

$IL = 10 * log_{10} \alpha = 0.048$ dB

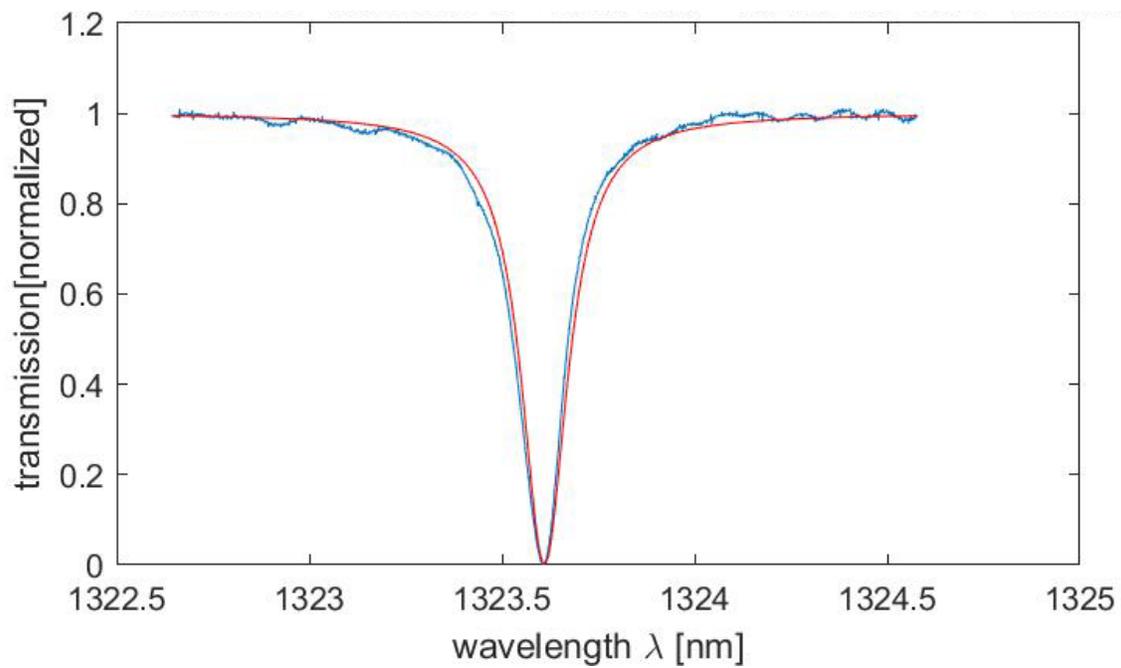

**Fig. S1.**
Transmission spectrum for a resonance of the memresonator including the Lorenztian fitted curve.

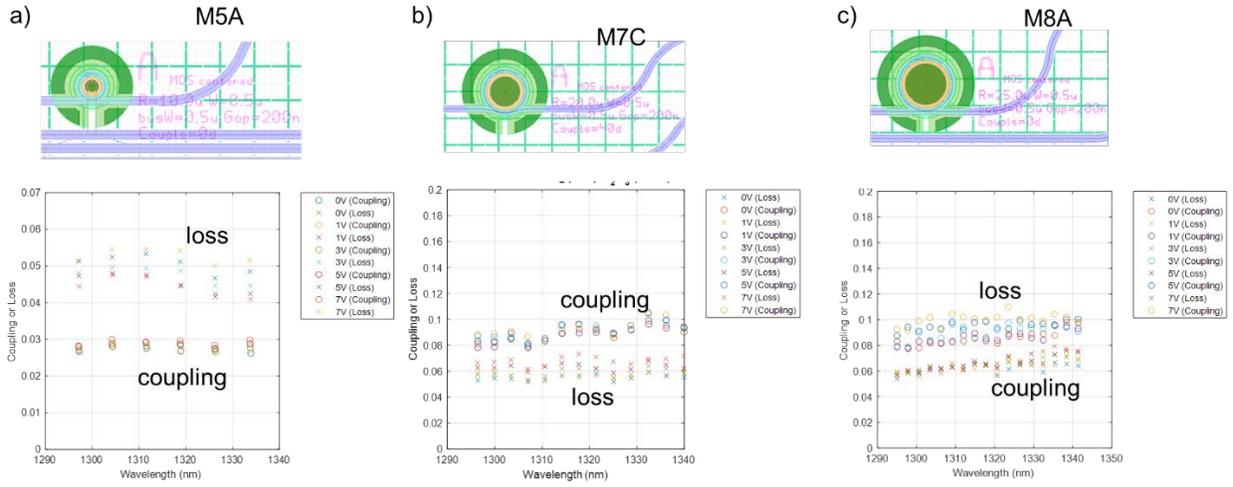

**Fig. S2.**

a) Insertion and coupling loss measured at varying applied bias voltages for a 10 μm radius memresonator, b) a 20 μm radius memresonator, c) and a 25 μm radius memresonator.

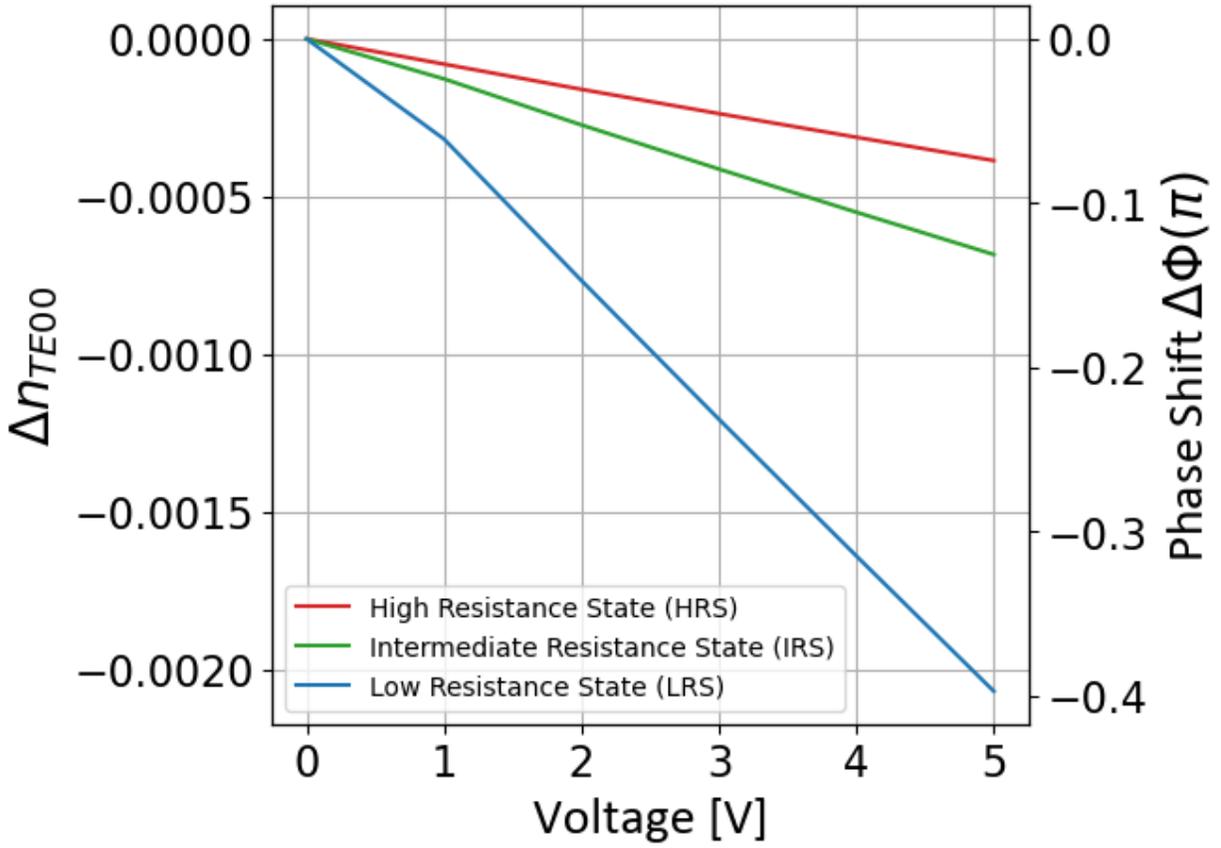

**Fig. S3.**
Effective refractive index change and phase shift as a function of bias voltage of the memresonator.

S3. Leakage Current

The leakage current of the memristor was measured in the high resistance state (HRS) and displayed in Fig. S6 using a voltage sweep from 0 to 5 V.

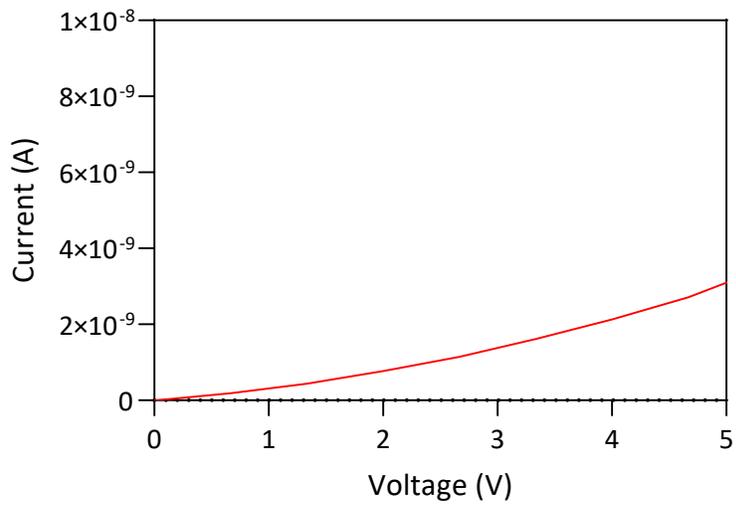

**Fig. S4.**
Leakage current of the memristor in the HRS.

## S4. Broadband Spectrum

We measured the broadband spectrum of the memresonator without extracting any losses from the grating couplers or the waveguide losses and plotted it in Fig. S7.

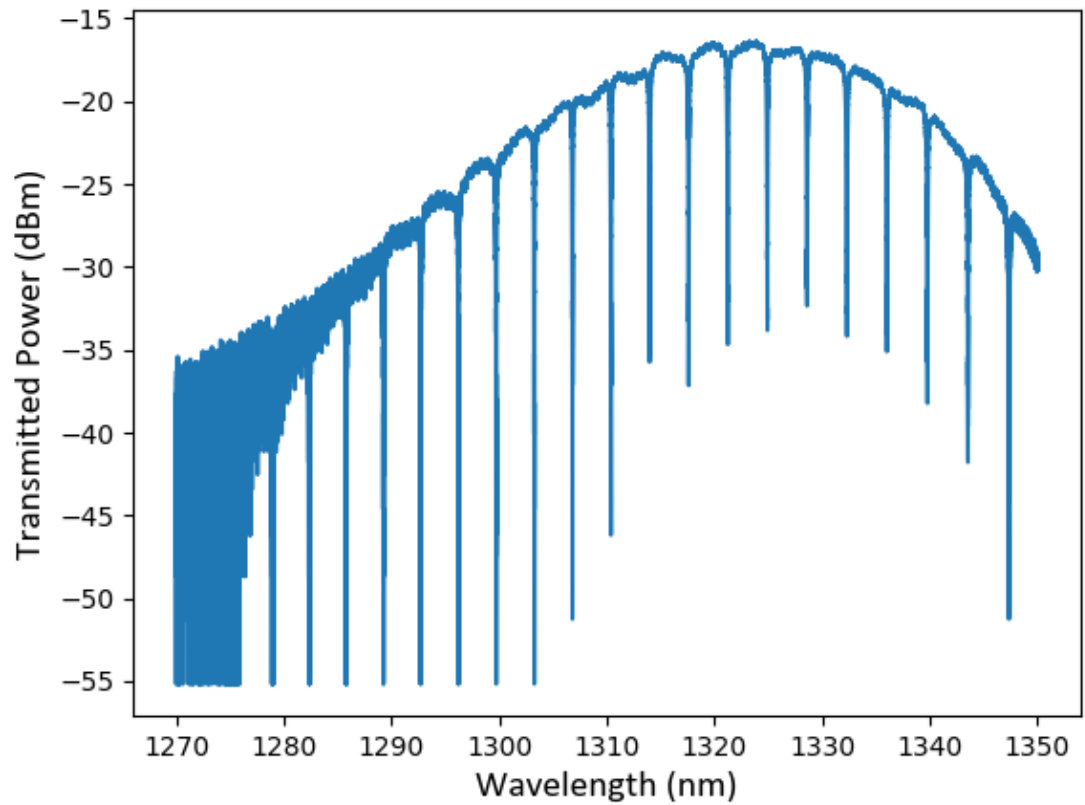

**Fig. S5.**
Broadband spectrum of the memresonator without extracting losses from the grating coupler.

S5. Phase Shift Calculation

In order to calculate the phase shift of the memresonator, we used the following formula:

$$Phase\ shift = \frac{2\pi\Delta\lambda}{FSR} = \frac{2\pi * 0.08\ nm}{2.808\ nm} \sim 0.18\pi$$

where FSR = 2.808nm, and the $\Delta\lambda$ = 0.08 nm between the LRS and the HRS at a read voltage of 2 V. This gives us an estimated $L_\pi$ of ($\pi$/0.18$\pi$) * 2*$\pi$*10µm = 0.349 mm ≈ 0.35 $mm$.

S6. Thermal simulations

A 15 nm × 15 nm × 20 nm filament of aluminum was inserted in the $Al_2O_3$ layer to act as a conductive path for current to flow in the charge, heat, and optical simulations. The simulations were performed using Ansys Lumerical Finite Element IDE and Finite Difference IDE. The dimensions of the aluminum conductive filament were chosen to replicate conductive filaments formed in similar devices [S1, S2, S3]. The charge simulations were done first to get the carrier concentration within the waveguide. The charge simulation data was then imported into the Finite Difference IDE solver to calculate the effective index of refraction for the TE modes in the waveguide. Thermal simulations were also done in Finite Element IDE and also imported into the Finite Difference IDE solver in the calculations of the effective index of refraction.

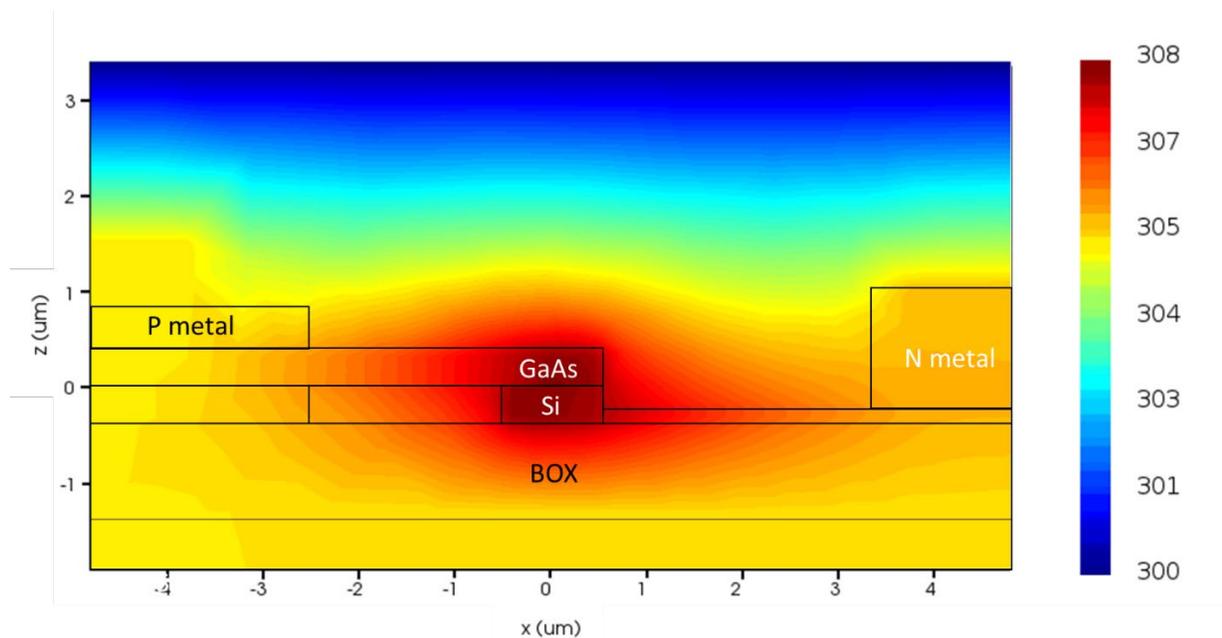

**Fig. S6.**
Thermal simulations of memristive III-V/Si waveguide in the LRS and 10 V applied. Localized Joule heating occurs within the CF located inside of the interfacial oxide layer between the III-V and Si.

**Supplementary Text References**

[S1] W. Sun, "Understanding memristive switching via in situ characterization and device modeling," *Nature Communications,* vol. 10, p. 13, 2019.
[S2] X. Zhang, L. Xu, H. Zhang, J. Liu, D. Tan, L. Chen, Z. Ma and W. Li, "Effect of Joule Heating on Resistive Switching Characteristic in AlOx Cells Made by Thermal Oxidation Formation," *Nanoscale Research Letters,* vol. 15, p. 11, January 2020.
[S3] Y. Wu, S. Yu, B. Lee and P. Wong, "Low-power TiN/$Al_2O_3$/Pt resistive switching device with sub-20 μA switching current and gradual resistance modulation," Journal of Applied Physics, vol. 110, p. 094104, November 2011.